\documentclass{aa}
\usepackage{txfonts}
\usepackage{graphicx}
\usepackage{natbib}
\bibpunct{(}{)}{;}{a}{}{,} 
\newcommand{\gr}{$\gamma$-ray \,}
\newcommand{\grs}{$\gamma$-rays \,}

\begin{document}

\title {Gamma-ray emission expected from Kepler's SNR}

   \subtitle{}

  \author{E.G.Berezhko
          \inst{1}
          \and
           L.T.Ksenofontov
          \inst{1}
          \and
          H.J.V\"olk
          \inst{2}
             }            

   \offprints{H.J.V\"olk}

   \institute{Yu.G.Shafer Institute of Cosmophysical Research and Aeronomy,
                     31 Lenin Ave., 677980 Yakutsk, Russia\\
              \email{berezhko@ikfia.ysn.ru}
         \and
             Max-Planck-Institut f\"ur Kernphysik,
                Postfach 103980, D-69029 Heidelberg, Germany\\
              \email{Heinrich.Voelk@mpi-hd.mpg.de}             
            }

   \date{Received month day, year; accepted month day, year}

\abstract 
{}
  {Nonlinear kinetic theory of cosmic ray (CR)
acceleration in supernova remnants (SNRs) is used to
investigate the properties of Kepler's SNR and, in particular, to predict
the \gr spectrum expected from this SNR.}
  {Observations of the nonthermal
radio and X-ray emission spectra as well as theoretical
constraints for the total supernova (SN) explosion energy
$E_\mathrm{sn}$ are used to constrain the astronomical and
particle acceleration parameters of the system.} 
  {Under the assumption that
Kepler's SN is a type Ia SN we determine for any given explosion energy
$E_\mathrm{sn}$ and source distance $d$ the mass density of the ambient
interstellar medium (ISM) from a fit to the observed SNR size and
expansion speed. This makes it possible to make predictions
for the expected \gr flux.  Exploring the expected distance range we find
that for a typical explosion energy $E_\mathrm{sn}=10^{51}$~erg the
expected energy flux of TeV \grs varies from $2\times 10^{-11}$ to
$10^{-13}$~erg/(cm$^2$s) when the distance changes from $d=3.4$~kpc to
7~kpc. In all cases the \gr emission is dominated by $\pi^0$-decay \grs
due to nuclear CRs. Therefore Kepler's SNR represents a very promising
target for instruments like H.E.S.S.,
CANGAROO and GLAST. A non-detection of \grs would mean that
the actual source distance is larger than 7~kpc.}
{} 

\titlerunning{Gamma-ray emission expected from Kepler's SNR}
    \maketitle

%

\section{Introduction}

Kepler's supernova remnant (SNR) (G4.5+6.8) is the result of a bright supernova
(SN) in our Galaxy that exploded in 1604. This SNR has been extensively
observed throughout the electromagnetic spectrum \citep[for a recent review,
see][and references therein]{blair05}. At the same time the type of Kepler's SN
has been debated over the years. Initially it was considered a type Ia SN, based on a study of the historical light curve of the
SN \citep{baade43}. In addition its location well above the Galactic plane
would be unexpected for a massive progenitor star.  More recently it was argued
that the light curve does not contradict a type II-L SN \citep{doggett85}, and
\cite{bandiera87} proposed a bow-shock model in which a massive star, ejected
from the Galactic plane, exploded into its own circumstellar medium. According
to this picture the remnant now interacts with the dense bow-shock shell that
was produced in the interaction of the progenitor's stellar wind with the
interstellar medium. Subsequently \cite{bork92} calculated a detailed model of
the SNR dynamics. It was argued that the observed morphology is in good
agreement with the bow-shock model. However, the thermal X-ray spectra,
obtained more recently with ASCA \citep{kinu99}, Chandra
\citep{hwang00} and XMM \citep{cassam04}, and corresponding theoretical
modeling \citep{badenes05}, favor a type Ia event. We take this as our
starting point.

The similarity of their light curves and spectra favor the idea that SNe of
type Ia form a rather homogeneous class with small variations in absolute
brightness. This leads to the assumption that the corresponding mechanical
explosion energy $E_\mathrm{sn}$ has also a very narrow distribution.  Despite
the considerable progress in modeling of type Ia SN explosions, up to now there
is no consensus about the mean value of the explosion energy.  Within the
so-called delayed-detonation model a typical range
$E_\mathrm{sn}=(1.3-1.6)\times10^{51}$~erg was obtained \citep{gamezo05}. This
agrees with the range obtained from several objects \citep{wheeler95}.

The deflagration model has resulted in considerably lower mean energy releases
$E_\mathrm{sn}=(0.4-0.6)\times10^{51}$~erg \citep{khokhlov00,reinecke02}. The
latter calculations however appear to underestimate to some extent the value of
the explosion energy \citep{reinecke02, hill04}, which is now expected to be
rather in the range $E_\mathrm{sn}=(0.8-1)\times10^{51}$~erg in this approach
(Hillebrandt, private communication). In this situation we use below the value
$E_\mathrm{sn}=10^{51}$~erg as a typical explosion energy for type Ia events.
Since the value of $E_\mathrm{sn}$ strongly influences the SNR dynamics and in
particular the expected \gr flux, we explore the range
$E_\mathrm{sn}=(0.5-2)\times10^{51}$~erg, in order to demonstrate the
sensitivity of the final results to the value of $E_\mathrm{sn}$.

The most recent radio study of the distance to the SNR by \cite{rg99} leads to
a lower limit of $4.8\pm 1.4$~kpc and an upper limit of 6.4~kpc. Therefore we
explore below the range $d=3.4-7$~kpc. For any given pair of values
$E_\mathrm{sn}$ and $d$ we find the density of the ambient interstellar medium
(ISM) from a fit to the observed SNR size and expansion speed \citep{dick88}.
This makes it possible to make quite definite predictions for the cosmic ray
(CR) and \gr production in this SNR. We apply here the
nonlinear kinetic theory of CR acceleration in SNRs \citep{bek96, bv97}. As was
successfully done for the remnants SN~1006, Cas~A and Tycho's SNR
\citep{bkv02,bkv03,bpv03,vbkr02,vbk05,bv04cas}, we use observations of the
nonthermal radio and X-ray emission spectra to constrain the astronomical
parameters as well as the particle acceleration parameters of the system, such
as the interior magnetic field strength and the CR injection rates. We show
that in all the cases considered the expected \gr flux is at a detectable level
if the source distance is not larger than 7~kpc. Therefore the detection of TeV
\grs from Kepler's SNR (that is expected in the near future) will enable us to
determine the SN explosion energy and the source distance.

\section{Model}
Hydrodynamically, a SN explosion ejects a shell of matter with total energy
$E_\mathrm{sn}$ and mass $M_\mathrm{ej}$. During an initial period the shell
material has a broad distribution in velocity $v$. The fastest part of this
ejecta distribution is described by a power law $dM_\mathrm{ej}/dv\propto
v^{2-k}$, following the treatments by \citet{jones81} and \citet{chev82}. The
interaction of the ejecta with the ISM -- assumed to be uniform -- creates a
strong shock which heats the thermal gas and accelerates particles
diffusively to a nonthermal CR component of comparable energy density. For
SN~Ia explosions we use the values $M_\mathrm{ej}=1.4M_{\odot}$ and $k=7$. Note that since Kepler's SNR
is already in the Sedov evolutionary phase its dynamics are not very sensitive
to the ejecta parameters $M_\mathrm{ej}$ and $k$.

Our nonlinear kinetic theory for this process is based on a fully
time-dependent, spherically symmetric solution of the CR transport equations,
coupled nonlinearly to the gas dynamic equations for the thermal gas
component. Since all relevant equations, initial and boundary conditions for
this model have already been described in detail in the above papers, we do not
present them here and only briefly discuss the most important aspects below.

The coupling between the ionized thermal gas (plasma) and the energetic
particles occurs primarily through magnetic field fluctuations carried by the
plasma which scatter energetic particles in pitch angle. In shock waves
this allows the diffusive acceleration of particles. The plasma physics of
the field fluctuations is not yet worked out in full detail.  Nevertheless, the
accelerating CRs strongly excite magnetic fluctuations upstream of the outer SN
shock in the form of Alfv\'en waves \citep[e.g.][]{bell78,blandford78} as a
result of the so-called streaming instability. In quasilinear approximation the
wave amplitudes $\delta B$ grow to very high values $\delta B > B$, where $B$
is the average field strength \citep{mckv82}. Since such fluctuations scatter
CRs extremely effectively, the CR diffusion coefficient can therefore be
assumed to be as small as its asymptotic limit $\kappa (p)=\kappa (mc)(p/mc)$,
the Bohm limit, where $\kappa(mc)=mc^3/(3eB)$, $e$ and $m$ are the particle
charge and mass, $p$ denotes the particle momentum, $B$ is the magnetic field
strength, and $c$ is the speed of light.

If $B_\mathrm{ISM}$ is the pre-existing field in the surrounding medium, then
the strong streaming instability would suggest that the instability growth is
restricted by nonlinear mechanisms to a level $\delta B\sim B_\mathrm{ISM}$,
any further turbulent energy being dissipated into the thermal gas
\citep{vmck81}.  Earlier attempts to give a full nonlinear description of the
magnetic field evolution through a numerical simulation \citep{lbell00,belll01}
concluded that a considerable amplification of the ``average'' magnetic field
in the smooth shock precursor -- produced by the finite CR pressure -- should
occur. It was expected that a non-negligible fraction of the
shock ram pressure $\rho_{0} V_\mathrm{s}^2$ ($\rho_{0}$ is the ISM density) is
converted into magnetic field energy. Subsequently, \citet{bell04} argued that
this amplification is the result of a nonresonant instability, giving rise to
the effective upstream field $B_0>B_\mathrm{ISM}$, on top of
which the Alfv\'en waves grow to amplitudes $\delta B\sim B_0$. The Bohm limit
and the gas heating are then to be calculated with $B=B_0$.

In our analyses of the synchrotron spectrum of SN 1006, Tycho's SNR, and
Cas~A, such magnetic field amplifications were indeed found; they can only be
produced as a nonlinear effect by a very effectively accelerated nuclear
CR component. Its energy density, consistent with all existing data, is so
high that it is able to strongly excite magnetohydrodynamic fluctuations, and
thus to amplify the upstream magnetic field $B_\mathrm{ISM}$ to an effective
field $B_{0}>B_\mathrm{ISM}$, at the same time permitting efficient CR
scattering on all scales, reaching the Bohm limit. The same large effective
magnetic field turns out to be required, within the errors, by the comparison
of this self-consistent theory with the field strength derived from the
morphology of the observed X-ray synchrotron emission, in particular its
spatial fine structure. We therefore also allow for the possibility
of an amplified field and assume the corresponding Bohm limit for the
particle diffusion coefficient.

We describe the number of suprathermal protons injected into the acceleration
process by a dimensionless injection parameter $\eta \ll 1$ which is a fixed
fraction of the gas particles entering the shock front. For simplicity we
assume that the injected particles have a velocity four times higher than the
postshock sound speed \citep[see][for a review]{md01}.  It is expected that ion
injection is quite efficient at the quasiparallel portion of the shock surface,
where it is characterized by values $\eta=10^{-4}$~to $10^{-3}$ \citep[see][for
details]{vbk03}.  Since this injection is expected to be strongly suppressed at
the quasiperpendicular parts of the shock, one should renormalize the results
for the nucleonic spectrum which were calculated within the spherically
symmetric model. The deviation from spherical symmetry, introduced by the
divergence-free nature of the magnetic field in the actual SNR, can be
approximately taken into account by a renormalization factor $f_\mathrm{re}<1$,
roughly $f_\mathrm{re}=0.15$~to $0.25$, which diminishes the nucleonic CR
production efficiency calculated in the spherical model, and all effects
associated with it.

We assume also that electrons are injected into the acceleration process at the
shock front. Formally their injection momentum is taken to be the same as that
of the protons. Since the physical processes of electron
injection are still poorly known, we choose the electron injection rate such
that the electron:proton ratio $K_\mathrm{ep}$ (which we define as the ratio of
their distribution functions $f_\mathrm{e}(p)/f(p)$ at all rigidities where the
protons are already relativistic and the electrons have not been yet cooled
radiatively) is a constant to be determined from the synchrotron
observations.

The electron distribution function $f_\mathrm{e}(p)$ deviates only at
sufficiently large momenta from the relation $f_\mathrm{e}(p)= K_\mathrm{ep}
f(p)$ as a result of synchrotron losses, which are taken into account by
supplementing the ordinary diffusive transport equation by a radiative loss
term.

From the point of view of injection/acceleration theory, we must treat
$K_\mathrm{ep}$, together with $B_0$ and $\eta$, as a theoretically until now
not very precisely calculable parameter, to be quantitatively determined
by comparison with the synchrotron observations \citep[see][for
reviews]{vlk03,ber05}

The solution of the dynamic equations at each instant of time yields the CR
spectrum and the spatial distributions of CRs and thermal gas. This allows
the calculation of the expected fluxes of nonthermal emission produced by the
accelerated CRs.

\section{Results and discussion}
\begin{figure}[t]
  \resizebox{\hsize}{!}{\includegraphics{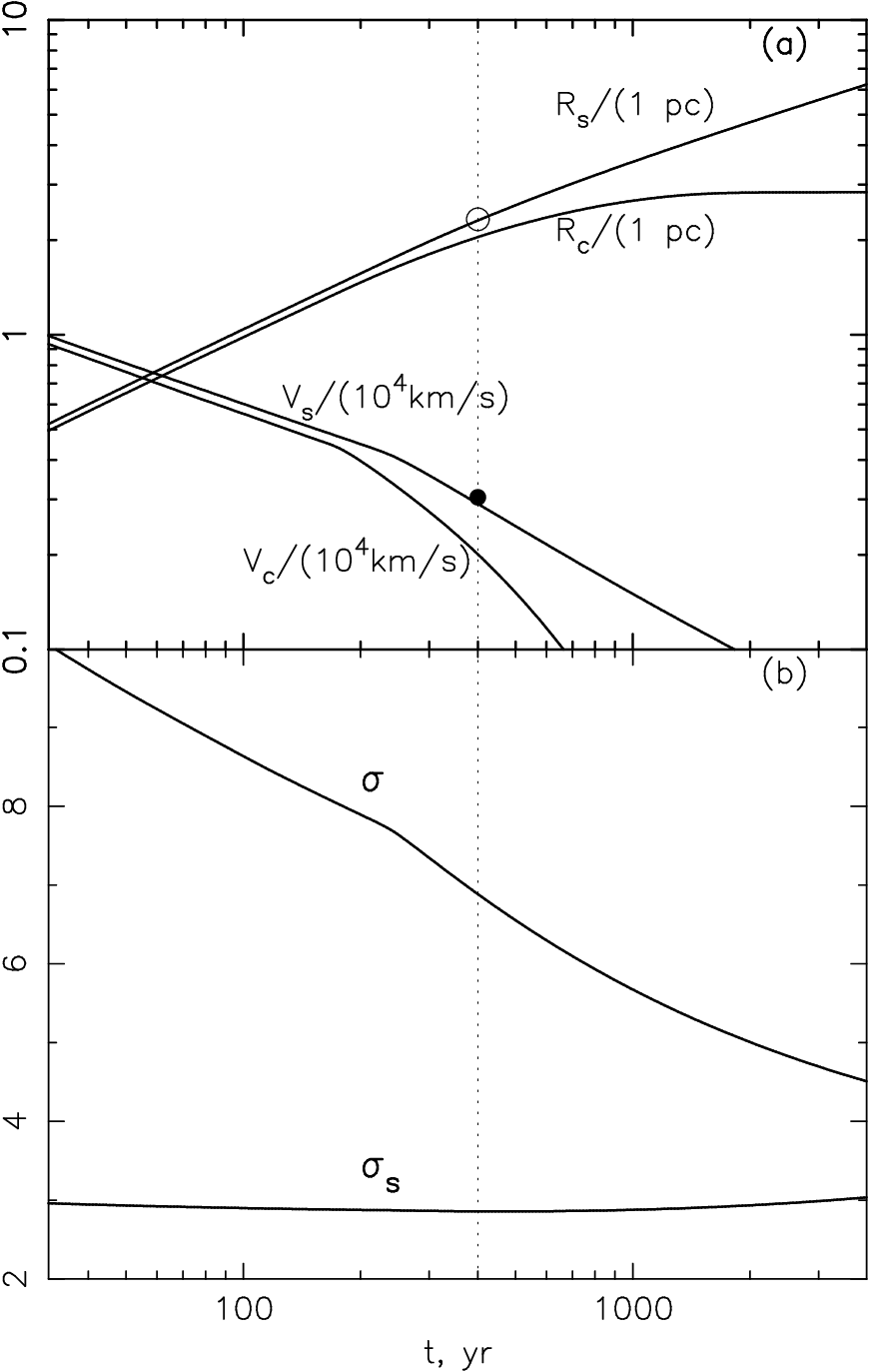}}
\caption{(a) Shock radius $R_\mathrm{s}$, contact
  discontinuity radius $R_\mathrm{c}$, shock speed $V_\mathrm{s}$, and contact
  discontinuity speed $V_\mathrm{c}$, as functions of time since explosion; (b)
  total shock ($\sigma$) and subshock ($\sigma_\mathrm{s}$) compression ratios.
  The dotted vertical line marks the current epoch $t_\mathrm{c}$.  The
  observed mean size and speed of the shock, as determined by radio
  measurements \citep{dick88}, are shown as well.}
\label{fig1}
\end{figure}

\begin{figure}[t]
\resizebox{\hsize}{!}{\includegraphics{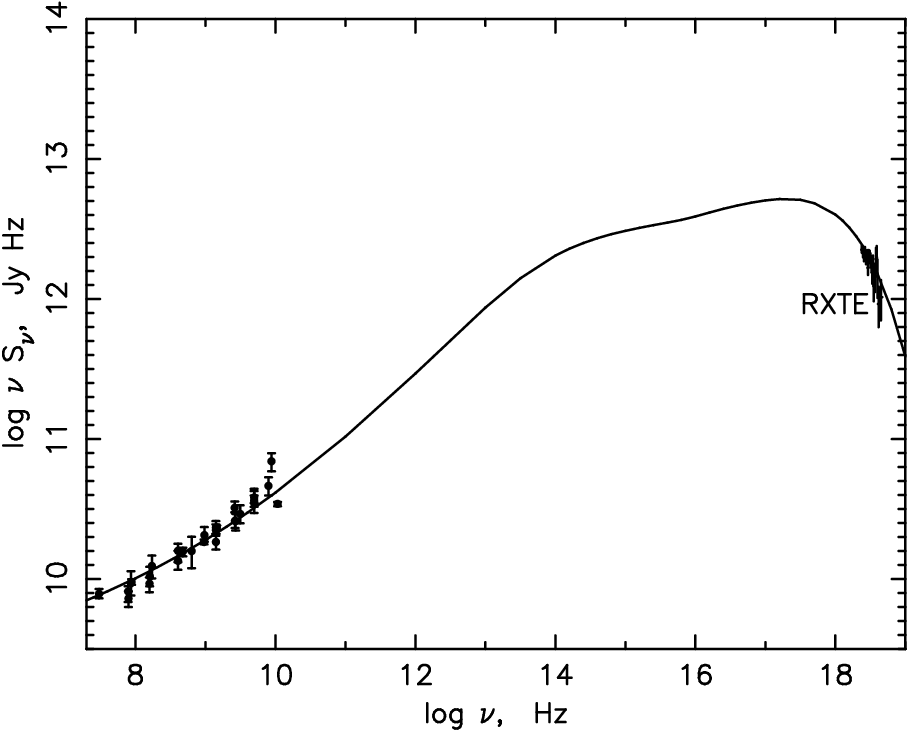}}
\caption{Calculated energy flux of synchrotron emission as a 
function of frequency for the
same case as in Fig.~\ref{fig1}. The observed non-thermal X-ray \cite{allen99}
and radio emission \cite{re92} flux values are also shown.} 
\label{fig2} 
\end{figure}

As a starting case we consider the typical value $E_\mathrm{sn}=10^{51}$~erg
for the explosion energy and the conventional distance $d=4.8$~kpc. The results
of our calculations are presented in Figs.~\ref{fig1}--\ref{fig2}. The hydrogen
number density $N_\mathrm{H}=3$~cm$^{-3}$, which determines the ISM density
$\rho_0=1.4m_\mathrm{p}N_\mathrm{H}$, was chosen to fit the size $R_\mathrm{s}$
and the expansion speed $V_\mathrm{s}$ at the present age
$t_\mathrm{c}=400$~yr. Such a value of $N_\mathrm{H}$ is clearly a spatial
average over the actual density structure. Optical $\mathrm{H}\alpha +
[\mathrm{N}~\mathrm{II}]$ data show knots and filaments, also influenced by
variable foreground extinction \citep[e.g.][for a review]{blair05}. For the
acceleration of the very high energy particles which give rise to the hard
X-ray synchrotron and TeV \gr emissions such small-scale inhomogeneities,
however, play essentially no role.

As shown in Fig.~\ref{fig1}a the current evolutionary phase of Kepler's SNR
corresponds to the Sedov phase. The adopted proton injection rate
$\eta=1.5\times 10^{-3}$ leads to a significant shock modification, characterized by a
total shock compression ratio $\sigma\approx 6.9$ and a subshock compression
ratio $\sigma_\mathrm{s}\approx 2.9$ (see Fig.~\ref{fig1}b). Such a shock
modification is needed to fit the observed steep radio spectrum and the smooth
connection with its X-ray part (see below).

About 10\% of the explosion energy has been transfered into CR energy up to
now, which means that the CR energy content is
$E_\mathrm{c}=0.1E_\mathrm{sn}$.

The calculated synchrotron flux is shown in Fig.~\ref{fig2} together with the
observed values at radio and X-ray frequencies. At radio frequencies the
synchrotron spectrum $S_{\nu}\propto \nu^{-\alpha}$ has spectral index
$\alpha=0.71$ \citep{delan02}. It deviates significantly from the value
$\alpha=0.5$ that corresponds to an unmodified strong shock. The adopted proton
injection rate $\eta=1.5\times 10^{-3}$ gives the required shock modification.
The elec\-tron-to-proton ratio $K_\mathrm{ep}=1.3\times 10^{-4}$ and an interior
magnetic field strength $B_\mathrm{d}=480$~$\mu$G give a good fit for the
experimental data in the radio and X-ray ranges.
Note that the interior magnetic field $B_\mathrm{d}=480$~$\mu$G, derived here
from the fit of the overall synchrotron spectrum, is higher than the
value $B_\mathrm{d}'=215$~$\mu$G, determined from the observed spatial fine
structure of the synchrotron emission \citep{vbk05}. However, if we use the
thinnest X-ray radial profile observed by Chandra \citep{bamba05}, which
has an angular width $\Delta \psi =2.1''$, we obtain $B_\mathrm{d}'= 340$~$\mu$G
in rough agreement with the value derived from the fit of the overall
synchrotron spectrum. Such a high interior magnetic field is the result of
field amplification by the nonlinear CR backreaction on the acceleration
process \citep[]{belll01, bell04}. It was recently established that such strong
field amplification takes place in all young Galactic SNRs which have known
filamentary structures in the nonthermal X-ray emission \citep{vbk05}.
%
%

\begin{figure}[t]
\resizebox{\hsize}{!}{\includegraphics{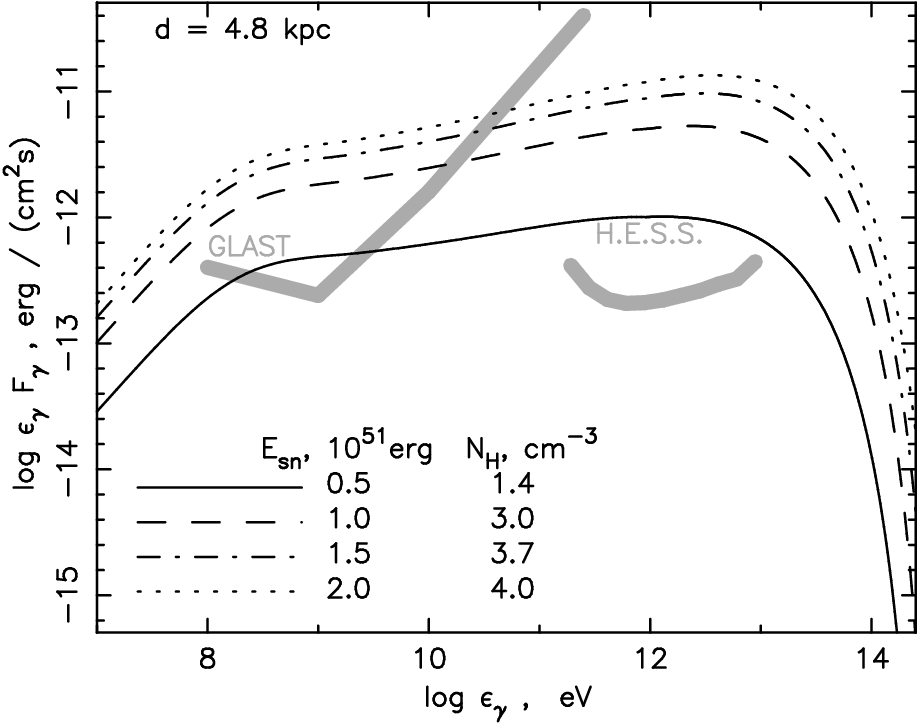}}
\caption{Total ($\pi^0$-decay + IC) integral $\gamma$-ray energy
  fluxes as a function of $\gamma$-ray energy for the source distance
  $d=4.8$~kpc and four values of the SN explosion energy
  $E_\mathrm{sn}/(10^{51}$~erg)=0.5, 1, 1.5, 2. For comparison, the respective
  sensitivities for a $5\sigma$ detection in one year with GLAST
  \citep{weekes03}, and in 50 hours at $20^{\circ}$ zenith angle for a photon
  index 2.6 (as for the Crab Nebula) with H.E.S.S. \citep{funk05}, are
  shown. }
\label{fig3}
\end{figure}

In Fig.~\ref{fig3} we present the gamma-ray spectrum of Kepler's SNR, expected
at the current epoch. It is mainly produced by the
CR proton component in hadronic collisions with background gas nuclei, leading
to $\pi^0$-production and subsequent decay into two gamma-quanta. This
so-called hadronic \gr component exceeds the leptonic \gr component due to the
Inverse Compton (IC) scattering off the cosmic microwave background by more
than a factor of $10^3$. The integral gamma-ray spectrum is expected to be very
hard, $F_{\gamma}\propto \epsilon_{\gamma}^{-0.8}$, within the energy range
from 1~GeV to almost 10~TeV. At $\epsilon_{\gamma}=1$~TeV $\epsilon_{\gamma}
F_{\gamma}\approx 5\times 10^{-12}$~erg/(cm$^2$s) for $E_\mathrm{sn} = 10^{51}$~erg.  Since the SN explosion
energy is not exactly known, we present in Fig.~\ref{fig3} also the results
calculated for the three other values $E_\mathrm{sn}/(10^{51}$~erg)=0.5, 1.5
and 2. We note that even at the lowest explosion energy
$E_\mathrm{sn}=0.5\times10^{51}$~erg considered here, the expected \gr flux
exceeds the sensitivity of the GLAST instrument at GeV energies and of
the H.E.S.S. instrument at TeV energies.  At TeV-energies the expected
energy flux is $\epsilon_{\gamma} F_{\gamma}\approx 10^{-12}$~erg/(cm$^2$s) in
the case $E_\mathrm{sn}=0.5\times10^{51}$~erg and an order of magnitude higher
for $E_\mathrm{sn}=2\times10^{51}$~erg.
%

Qualitatively, the dependence of the expected \gr flux on the explosion
energy can be understood if one takes into account that the three relevant
parameters -- distance $d$, explosion energy $E_\mathrm{sn}$ and ISM density
$N_\mathrm{H}$ -- are connected by the relation
\begin{equation}
N_\mathrm{H}\propto E_\mathrm{sn}/d^5,
\label{eq1}
\end{equation}
because in the Sedov phase the SNR radius $R_\mathrm{s}\propto d$ is determined
by the relation $R_\mathrm{s}\propto
(E_\mathrm{sn}/N_\mathrm{H})^{1/5}t^{2/5}$.  Taking into account the dependence
of the $\pi^0$-decay \gr flux on the relevant parameters
\begin{equation}
F_{\gamma} \propto E_\mathrm{c} N_\mathrm{H}/d^2,
\label{eq2}
\end{equation}
together with the fact that in the Sedov phase the ratio $E_\mathrm{c}/E_\mathrm{sn}$ is a weakly
dependent function of time, we have
\begin{equation}
F_{\gamma} \propto E_\mathrm{sn}^2/d^7.
\label{eq3}
\end{equation}
For a given source distance $d$ this gives $F_{\gamma}\propto E_\mathrm{sn}^2$,
in rough agreement with the numerical results presented in
Fig.~\ref{fig3}.

Note that magnetic field strengths vary very little with the explosion
energy: for the interval $E_\mathrm{sn}=(0.5-2)\times 10^{51}$~erg considered,
we have $B_\mathrm{d}=440-500$~$\mu$G and $B_\mathrm{d}'=346-337$~$\mu$G,
whereas the electron:proton ratio $K_\mathrm{ep}=(2.8-0.7)\times 10^{-4}$
varies roughly inversely proportional to the energy $E_\mathrm{sn}$.

\begin{figure}
\resizebox{\hsize}{!}{\includegraphics{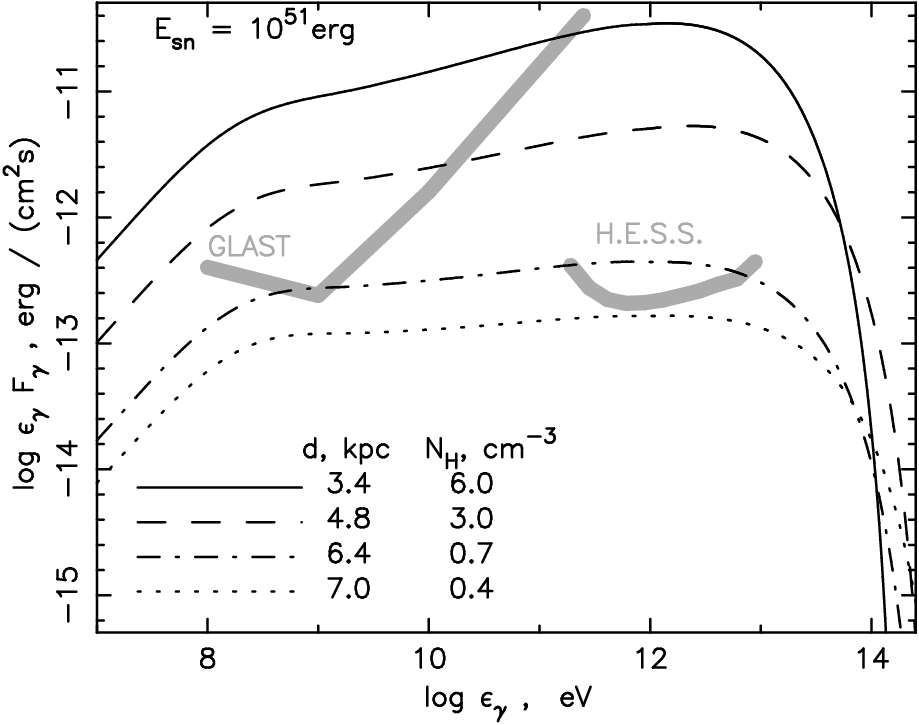}}
\caption{Same as in Fig.~\ref{fig3}, but different curves correspond now to
different source distances $d=3.4$, 4.8, 6.4, 7~kpc for the explosion energy
$10^{51}$ erg.}
\label{fig4}
\end{figure}

Since the source distance is not known very well, we performed our calculations
for a range of distances $d=3.4-7$~kpc in a similar way as it was done above
for $d=4.8$~kpc. In each case we achieve the same quality of fit of the
observed SNR size, its expansion speed and the overall synchrotron emission
spectrum. Therefore we present in Fig.~\ref{fig4} only the results of the \gr
energy fluxes expected for the SN explosion energy $E_\mathrm{sn}=10^{51}$~erg
and for four different distances from the range $d=3.4 - 7$~kpc. It can be seen
from Fig.~\ref{fig4} that Kepler's SNR is expected to be as bright a TeV \gr
source as the Crab Nebula if the distance is as small as $d=3.4$~kpc. The
expected \gr flux goes down with increasing distance and comes to the minimum
observable H.E.S.S. flux if the distance becomes as large as 7~kpc.

According to Eq.~(\ref{eq2}), for a fixed value of the SN explosion energy
$E_\mathrm{sn}$ the expected \gr flux is
\begin{equation}
F_{\gamma}\propto d^{-7}
\label{eq4}
\end{equation}
in good agreement with the results presented in Fig.~\ref{fig4}.

We note here that according to the so-called delayed detonation model
\citep{gamezo05} the value $E_\mathrm{sn}=10^{51}$~erg is even below the
lower end of the expected range of type Ia SN explosion energies. From this
point of view the results presented in Fig.~\ref{fig4} most probably give a
lower limit for the expected \gr flux for any given distance $d$, and therefore
these results convince us that Kepler's SNR is a potential high-energy \gr
source.
%

We emphasize that an increase of the assumed source distance leads to an
increase of the linear width $L\propto d\Delta \psi$ of the filaments observed
in nonthermal X-rays. This leads in turn to a decrease of the interior magnetic
field strength $B_\mathrm{d}'\propto L^{-2/3}$, determined from the filament
size. On the other hand, the linear shock size $R_\mathrm{s}\propto d$ and the
shock speed $V_\mathrm{s}\propto d$ become larger for larger $d$ and this makes
the CR acceleration process faster.  Therefore a higher magnetic field
$B_\mathrm{d}$ with a correspondingly more intense synchrotron cooling effect
of CR electrons is required in order to provide a smooth cutoff of the
synchrotron spectrum $S_{\nu}(\nu)$ consistent with the existing nonthermal
X-ray measurements (see Fig.~\ref{fig2}). As a result, at $d=7$~kpc we have
$B_\mathrm{d}=534$~$\mu$G, whereas $B_\mathrm{d}'=295$~$\mu$G is already
considerably lower. This means that a source distance $d\approx 7$~kpc or
larger has to be considered as inappropriate if the radial X-ray profile width
$\Delta \psi=2.1''$ observed by Chandra \citep{bamba05} is indeed the
smallest one in this source. Then Kepler's SNR has to be considered as a
promising potential source of high-energy \grs. We note also that the
electron:proton ratio $K_\mathrm{ep}$ is quite insensitive to the distance $d$.

It is clear that the asymmetry of the X-ray emission is evidence that the
  surrounding ISM is inhomogeneous on a scale comparable with the SNR size
  \citep[e.g.][]{bandiera87}. Our spherically symmetrical approach is not able
  to take this into account. However, since we used the SNR size $R_\mathrm{s}$
  and speed $V_\mathrm{s}$ as averages over the remnant, we believe that the
  ISM density adopted for each pair of values $d$ and $E_\mathrm{sn}$ also
  represents the ambient density averaged over the remnant and thus furnishes a
  reliable estimate for the expected total gamma-ray emission.

\section{Summary}

Our consideration of CR acceleration and nonthermal emission in Kepler's SNR
demonstrates that a spherically symmetric nonlinear kinetic theory reproduces
the SNR dynamics and the properties of its nonthermal radiation in a very
satisfactory way when Kepler's SN is treated as a type Ia SN.

The \gr energy flux expected at TeV energies is $\epsilon_{\gamma}
F_{\gamma}\approx (3-5)\times 10^{-12}$~erg/(cm$^2$s) if the distance is as
small as $d=4.8$~kpc.  The flux is expected to be in a detectable range
$\epsilon_{\gamma} F_{\gamma}> 10^{-13}$~erg/(cm$^2$s) at TeV energies if the
distance does not exceed 7~kpc. If the upper limit for the source distance is
indeed $d=6.4$~kpc \citep{rg99} -- a conclusion that is confirmed by the
consistency check of the interior magnetic field values obtained by two
independent methods -- we conclude that Kepler's SNR is a potentially bright \gr source in the sky.


\begin{acknowledgements} This work has been supported in part by the
  Russian Foundation for Basic Research (grant 03-02-16524) and LSS 422.2003.2.
  EGB and LTK acknowledge the hospitality of the Max-Planck-Institut f\"ur
  Kernphysik, where part of this work was carried out. The authors thank
    W. Hillebrandt and G. P\"uhlhofer for discussions on type Ia SN explosions
  and the specific properties of Kepler's SN.

\end{acknowledgements}

\end{document}